# Public perspectives on the design of fusion energy facilities

Nathan Kawamoto, Daniel Hoover, Jonathan Xie, Jacob Walters, Katie Snyder, Aditi Verma*
aditive@umich.edu
University of Michigan

## Abstract

As fusion energy technologies approach demonstration and commercial deployment, understanding public perspectives on future fusion facilities will be critical for achieving social license, especially because fusion energy facilities, unlike large fission reactors, may be sited in closer proximity to communities, due to distinct regulatory frameworks.

In a departure from the ‘decide-announce-defend’ approach typically used to site energy infrastructure, we develop a participatory design methodology for collaboratively designing fusion energy facilities with prospective host communities. We present here our findings from a participatory design workshop that brought together 22 community participants and 34 engineering students. Our analysis of the textual and visual data from this workshop shows a range of design values and decision-making criteria with ‘integrity’ and ‘respect’ ranking highest among values and ‘economic benefits’ and ‘environmental protection/safety’ ranking highest among decision-making criteria. Salient design themes that emerge across the facility concepts include connecting the history and legacy of the community to the design of the facility, respect for nature, care for workers, transparency and access to the facility, and health and safety of the host community. Participants reported predominantly positive sentiments, expressing joy and surprise as the workshop progressed from learning about fusion to designing the hypothetical facility. Our findings suggest that carrying out participatory design in the early stages of technology development can invite and make concrete perspectives on public hopes and concerns, improve understanding of, and curiosity about, an emerging technology, build toward social license and inform context-specific development of future fusion energy facilities.

## Introduction

As fusion energy technologies rapidly approach demonstration and commercial deployment, technology developers are starting to evaluate and select sites for fusion power plants. Initial interest (by companies such as Commonwealth Fusion Systems [1] and Zap [2]) has focused on retiring coal plants, but a variety of other use cases resulting from energy demand drivers [3] (for powering data centers, industrial process heat, increasing electrification of transportation, energy needs in remote communities, and growing household demand in grid-connected communities) are likely to lead to the consideration of other sites.

Because fusion energy systems will be regulated differently than fission reactors [4], fusion developers expect that these systems can be built in much greater proximity to communities and population centers than fission facilities historically have been. The development of fusion energy technologies and their potential deployment at scale also comes at a time of energy infrastructure expansion, when already ubiquitous energy infrastructure is likely to become even more embedded in the built environment. Given the potential for integrating fusion energy facilities into and around towns and cities, and near public areas, we explore, in this project, what perspectives publics bring to the fusion energy facility design process. We report on the process and outcome of a community design workshop through which community members, working with engineering students, created designs for hypothetical fusion energy facilities.

Participants from the local community, across a range of ages and backgrounds, joined in a day-long in-person workshop to collaborate on the design of a hypothetical fusion energy facility in southeast Michigan. Using participatory design methodology, the workshop invited participants working in teams to share values, establish design criteria, and collaboratively design their facility. Analysis of the workshop data, collected from participant workbooks, written reflections, audio recordings, and AI-generated visualizations of fusion energy facility concepts, suggest that participatory design is a promising strategy for building social license in early-stage development of energy facility design. This process appears to be successful in making community members' hopes and concerns clear and visible, improve publics' understanding of emerging energy technology, and support mutual learning.

Concretely, we observed several shared as well as distinct values and decision-making criteria developed by the participants. Shared values that ranked the highest across teams included 'integrity' and 'respect' whereas 'economic benefits' and 'environmental protection/safety' ranked the highest among shared decision-making criteria across teams. Further, several salient design themes also emerged in the facility concepts developed by our workshop participants. These included integrating the history and legacy of the community in the design of the facility, a respect for nature, care for the workers, transparency and access to the facility, and health and safety of the community. Though it is possible that these themes may more broadly be representative of community priorities in other places as well, our workshop demonstrated that each team interpreted these themes in unique ways in the design of their hypothetical facility. The existence of both shared as well as unique values, decision-making criteria, and design choices suggests there is value in carrying out similar participatory design efforts on a large scale to understand similarities and variations in public preferences, hopes, and concerns, even in the early stage of fusion energy development. The discovery of shared themes and priorities can inform place invariant design features of fusion energy systems whereas unique and context-specific preferences can be accounted for in designing other aspects of a fusion energy facility for particular locations. Such early participatory design efforts can therefore help establish design and research priorities for fusion energy systems in alignment with public preferences.

In this paper we explain our rationale for doing this work, our methodology and findings, and their implications for the deployment of fusion energy.

## Fusion energy: socioeconomic and environmental impacts

A primary motivation for the development of fusion energy lies in its potential to be an economically viable, clean, long-term source of energy. Over 45 companies around the world are actively working to commercialize fusion energy technologies, with over three dozen of these companies being based in the US [5,6]. Given the growing demand for energy both in the West and the Global South [7], fusion energy technologies, if commercialized in a timely manner, are expected to create value in the trillions of dollars [8]. A significant portion of this value is expected to arise from the positive local socioeconomic impact of the future fusion industry through the creation of local manufacturing, supply chains, and research centers, and the workforce requirements across these activities. The emergence and location of existing fusion companies in the US already suggests the potential for the creation of local innovation hubs or clusters – with such clusters already emerging in the Pacific Northwest (Seattle), Madison, New Jersey, and Boston, with others likely to be created in the US and around the world.

The emergence of the fusion industry has in part been led and accompanied by fusion industry groups (such as the Fusion Industry Association in the US and J-Fusion in Japan) and concerted national policies that support the rapid commercialization of fusion energy. In the US, the White House announced a 'Bold Decadal Vision for Fusion Energy' accompanied by the launch of the Milestones program for fusion energy commercialization [9]. The Milestones program currently supports eight fusion energy companies. Elsewhere, the EU supports fusion energy commercialization through its EUROfusion program [10], the UK launched its Fusion Futures program in 2023 [11], Germany launched its Fusion 2040 program in 2024 [12], and Japan has a Moonshot program for fusion energy [13]. Russia, China, and India, too, though not having named fusion energy development programs, have also increased investments in fusion energy. By some estimations, these nationally led initiatives are expected to pave the way for commercial fusion energy on a timescale likely to eclipse the ITER initiative [14]. Each of these nationally funded initiatives signals an intent to accelerate the development and deployment of fusion energy with the aim of realizing commercial fusion energy systems by mid-century.

These concerted efforts to accelerate the deployment of fusion energy are also leading to a renewed emphasis on and analysis of the environmental impacts of fusion energy. A significant advantage of fusion energy systems emphasized by developers is that they are not expected to generate high-level radioactive waste. One concern, however, that does arise is the generation of significant quantities of low-level waste [15,16]. The safe management of this low-level waste will likely require the creation of new infrastructure, policies, and facilities, as current low-level waste facilities (designed and sized to receive low-level waste from fission plants) are unlikely to be sufficient. Still other societal and environmental concerns that have been raised about fusion energy by experts and critics relate to their proliferation potential or resistance, the reliance on critical minerals for the development of fusion energy systems, and the need to safely manage and store the large volumes of tritium that will be generated during the operation of fusion power plants [17–20]. Several solutions and assessments of each of these concerns have been proposed. Though one-size-fits-all approaches likely do not exist, these concerns will have to be addressed on a case-by-case basis for each type of fusion energy system. More localized concerns that are likely to be raised by host communities of fusion energy facilities will relate to land and water use, localized environmental and socioeconomic impacts, and decommissioning of the facility at the end of its life, as well as the safety of fusion energy facilities throughout their lifetime [19,21,22]. It is now widely accepted that successfully addressing these concerns broadly as well as in a localized way for potential host communities will be key for securing ‘social license’ for fusion energy technologies [23].

## Public perspectives on fusion energy technologies

Survey evidence emerging from Europe as well as the US suggests a positive public outlook toward fusion energy [24–26]. However, it should be noted that this public sentiment is largely based on positive associations with the technology and not yet grounded in an understanding of fusion energy technologies, as well as their potential positive and negative socioeconomic and environmental impacts. Simultaneously, recent developments in the fusion regulatory landscape suggest that fusion is likely to be regulated significantly differently from fission [4,27] – which will potentially enable both more accelerated review and approval timelines for new projects as well as reduced emergency and planning zones – both of which suggest that fusion energy projects, once initiated, could both proceed quite quickly from inception to realization and that they might also be sited in significant proximity to people and communities.

Our motivation for the work described in this paper arises from the potential for relatively accelerated timelines from project initiation to realization, the relatively slower current timelines for technology development, and the imperative to understand and address public concerns as a prerequisite for the successful use of fusion energy at scale.

## Nuclear fission technologies: A mixed but largely cautionary tale

A cautionary tale on how not to approach public engagement can be found in the experience of the nuclear fission industry. Taking its lead from the risk perception studies carried out in the 1960s [28], the nuclear fission industry embarked on extensive campaigns of public education. This largely one-way communication drew on risk metrics and comparisons of the riskiness of nuclear fission with other risky activities as a way to educate and convince ‘the public’ that nuclear fission was a safe technology. We use the word ‘publics’ here deliberately because indeed the public is not a monolithic whole. These unidirectional programs of public engagement were unfortunately coupled with an approach to siting nuclear fission facilities that was not consultative [29,30]. The approach, infamously known as the ‘decide-announce-defend’ model of decision-making, did just that. Local host communities were minimally or not at all consulted in the initial stages of facility siting and project development. Instead, the decision to site the facility was typically defended once key decisions were made, through public hearings, often at a point where public input could do little to alter key design or siting decisions. This approach was not unique to nuclear energy facilities designed and built from the 1960s to the 1990s, but the unique, high-technology, high-hazard, and complex nature of fission reactors led to a heightened sense of concern, antagonism, and even opposition in the US and around the world – leading in many cases to the rejection of projects after significant investments in their development, suspension of construction, or in extreme cases, the rejection of plant operation once the facility was completed [31–33].

Paradoxically, communities that have accepted and hosted nuclear facilities have come to identify with them as a key feature of the community and local economy, with several such communities even expressing regret and disappointment when the nuclear facility in question approaches closure and decommissioning [34,35]. At a recent Stakeholder Engagement conference held by the International Atomic Energy Agency in 2025, over three dozen mayors from around the world shared their perspectives on local nuclear energy facilities, asserting that these facilities are a central part of their communities [36]. Understanding of and familiarity with nuclear technology has even led communities to consider hosting not just energy but also waste facilities [37,38]. The paradox of public acceptance, then, is that the history of fission technology is littered with failures to effectively engage publics and host communities, but successful engagement and siting have led to nuclear infrastructure in many cases becoming a desired and central feature of a community. Engagement failures have resulted primarily from unidirectional communication that seeks to 'educate', rather than mutual learning that seeks to understand community priorities, hopes, and concerns, including how publics and communities perceive the risks and benefits of new technologies.

Unlike the early risk perception literature (that shaped the fission industry's engagement with the public) that framed risks mainly in terms of mortality risk, more recent findings suggest that the publics' framings of risk are more nuanced, account for a range of potential concerns beyond mortality, and also that these framings of risks can be shaped by affect and emotion [39–41]. These ways of framing risks are not irrational or flawed but, in fact, deeply human. This knowledge about how the publics perceive new technologies calls for a more nuanced approach to public engagement – one that confers greater agency and more degrees of freedom to publics. These approaches are likely to be more conducive to eliciting public perspectives and facilitating mutual learning between publics and experts.

## Involving the public in technology development: participatory design and assessment

Broadly, these bidirectional approaches to public engagement are consultative and directly involve members of the public or potential host communities in the technology assessment and design decisions. Examples of these approaches include citizen advisory boards [42], citizen juries [43], deliberative polling [44], living labs [45], participatory technology assessment [46], and participatory design [47]. Many of these approaches were pioneered in Scandinavian countries and originated from a desire to give people and workers greater control in designing workplace technologies. Since their inception, these approaches have been used widely across a range of settings for the design of physical artifacts, tangible and intangible systems – including consumer products, automobiles, medical devices, software, urban infrastructure, and most recently, energy facilities [47]. The specific approach we use in our work is participatory design. Our use of this approach is predicated on the very early-stage nature of fusion energy technologies. Being in an early stage of development, no fully formed designs exist that the public can review, assess, and react to. This is, in many ways, an advantage because it allows publics the freedom and opportunity to explore and understand the underlying scientific and engineering principles and propose facility-level designs that would work for their specific community, in alignment with their preferences, hopes, and concerns.

# Methods

In this paper, we share the results of a workshop in which community members and engineers worked together to design hypothetical fusion energy facilities. The workshop findings shed light not only on the preferences for fusion energy facility design as articulated by our workshop participants, but also have procedural implications, indicating that such participatory design efforts can be carried out even in very early stages of technology development – as a way to elicit public input and inform key design decisions. Our participatory design workshops with community members were held as part of a course on the community-engaged design of energy technologies that Verma and Snyder teach at the University of Michigan [48,49]. As part of the course, students learn the fundamentals of nuclear science and engineering, technical communication, and participatory design to develop a sociotechnical competence in future inventors and designers of nuclear technologies. As part of the course term project, students work with community members to design a hypothetical energy facility. In Fall 2023, our focus was on fusion energy facility design. These community engagements occurred in the form of two structured sessions – a virtual workshop held midway during the semester and an in-person workshop at the end of the semester. The virtual workshop was dedicated to questions of process, while at the in-person workshop, community members and students worked in teams to create their fusion energy facility designs. In this paper, we focus on our findings from the in-person facility design workshop. We begin by describing the workshop design and participant demographics, and then describe our data collection and analysis methods in the remainder of this session before turning to the results.

## Workshop design

The in-person workshop, held in December 2023, unfolded over the course of five hours, starting at 11:30 am EST. The workshop was held on Saturday so that community participants would not have to take time off to attend. Each workshop participant received an honorarium for their time and participation.
In preparation for the workshop, the thirty-four students who participated were divided into ten teams and assigned distinct roles and responsibilities. These included: **Facilitators** who were responsible for leading conversations at each table as directed by the instructors of the workshop; **Archivist**s who collected audio recordings at each table using a phone or similar device. Archivists also took pictures of the exercises and collaborative processes completed during the workshop; **Storytellers** who led the table in preparing a final presentation and presented it with community participants on their respective teams, and **participant-observers** who took notes based on table activities and discussions. For teams that did not have four students, the participant-observer role was omitted as a distinct role, and those responsibilities were shared among students.

Throughout the workshop, the online tool Mentimeter was used to provide participants with an anonymous way to share ideas with all participants. This tool allowed prompts to be displayed on the projected screen, as well as displaying answers. During the workshop, Mentimeter was used by all participants for the entire group to answer prompts and spark discussions between all the teams. Participants were also given a workbook that was designed specifically for the workshop. Workbook pages corresponded with the structure of the workshop described below. All participants (students and community members) could document their thoughts, ideas, reactions, hopes, or concerns in the workbook if they preferred not to speak them aloud.

The workshop was structured into distinct time blocks as follows:

**Welcome and framing**: Participants were introduced to the workshop, participatory design, and the community-engaged design course. The workshop facilitators (Verma and Snyder) introduced community guidelines that would be followed during the workshop. These included principles such as “welcoming diverse points of view” and “practicing curiosity.” Participants also shared their own guidelines and values on Mentimeter. Finally, participants introduced themselves to their design teams, sharing their name, expertise, and top three values, which were all collected in their workbooks.

**Introduction to fusion**: Participants were given a brief overview of fusion energy. This included information about what nuclear fusion is, how it produces energy, and the benefits and drawbacks. Community participants were then encouraged to discuss these concepts with engineering students within their groups to solidify their understanding.

**Divergent thinking**: Participants were provided with a brief presentation about divergent thinking [50] and speculative design [51]. After, participants were provided with paper, writing utensils, sticky notes, and an idea map. Participants used sticky notes to independently write down ideas for a fusion facility that fit into one of six categories: “Technical”, “Socioeconomic”, “Environmental”, “Community”, “Aesthetic”, and “Wildcard.” After brainstorming ideas, participants worked together to sort these ideas onto the idea map corresponding with the area of the facility where the idea fit best. The design of this portion of the workshop was informed by a protocol developed for the subject matter expert design workshop [52].

**Decision thinking:** Participants were given a brief presentation on decision-making criteria. Using what they learned, participants brainstormed criteria on sticky notes. Teams ideated criteria such as affordability and safety. After organizing their criteria, participants were provided with stickers to put onto sticky notes as votes. For this exercise, each participant voted on five criteria and identified their top 3. Participants used these criteria to vote on the ideas that they had brainstormed in the divergent thinking exercise. There was no limit to how many votes each participant could use for their facility ideas. Participants were also asked to reflect on how it felt to work in their group. How did it make them feel when their idea was selected? What if it was not selected?

**Prototyping and Presentations**: Participants used AI image generators to visualize their fusion energy facility designs. Each group’s storyteller led the creation of group presentations and worked with a community participant to present their group’s work. Participants were encouraged to take notes on presentations in their workbooks.

**Wrap-Up**: Participants shared two words that they would use to describe the workshop. They were also given closing information during this time.

## Participant recruitment and demographics

The workshop participants included 22 community members and 34 students. Community participants were recruited through an open participation call and dissemination of information on the workshop that included posting of physical fliers around Ann Arbor, as well as digital fliers on a wide range of Facebook groups focused on Southeast Michigan, community forums such as Nextdoor, and email lists.

In this class, students were placed into 10 teams consisting of 3 to 4 students. It should be noted that these groups were not randomly generated. Each student filled out a survey regarding how they worked collaboratively and their study habits. The instructors used this information to form groups between people that they felt would work well together and complement one another's skills and interests. During the workshop, community participants were randomly added to these groups to create the final workshop groups. Tables 1 and 2 below show the participant demographics. Students and participants worked together in teams over the course of the workshop, with ten teams working in parallel.

Table 1. Demographics of workshop participants (community members)

| **Racial Demographics** | |
|---|---|
| White | 11 |
| Asian | 7 |
| Black | 1 |
| Hispanic or Latino | 1 |
| Multiracial | 1 |
| South Asian | 1 |

Table 2. Education and Employment (community members)

| **Education and Employment** | |
|---|---|
| Graduate/Professional Degree | 13 |
| Some College | 4 |
| High School Diploma | 2 |
| Students | 6 |
| Employed | 16 |

## Data collection and analysis

A variety of qualitative and quantitative data were collected during the workshop and interviews, including recorded discussions, photos of collaborative maps, and completed workbooks, which were later digitized verbatim. Our analysis aims to assess participant sentiment and how it evolved, as well as quantify community values through votes, idea frequency, and categories from the maps and workbooks.

The maps were digitized into a machine-readable format, with ideas, map regions, categories, and votes recorded. These ideas were then manually categorized using a joint clustering procedure with human coders and a large language model (ChatGPT 4.0) used for intercoder reliability (Fig. 1). Initially, ideas with similar meanings were grouped into clusters of varying sizes, then refined into 7 general categories. Cross-coder reliability was established by re-sorting ideas using ChatGPT, and any discrepancies between human and LLM categorizations were resolved by the research team. Figure 1 below describes this process.

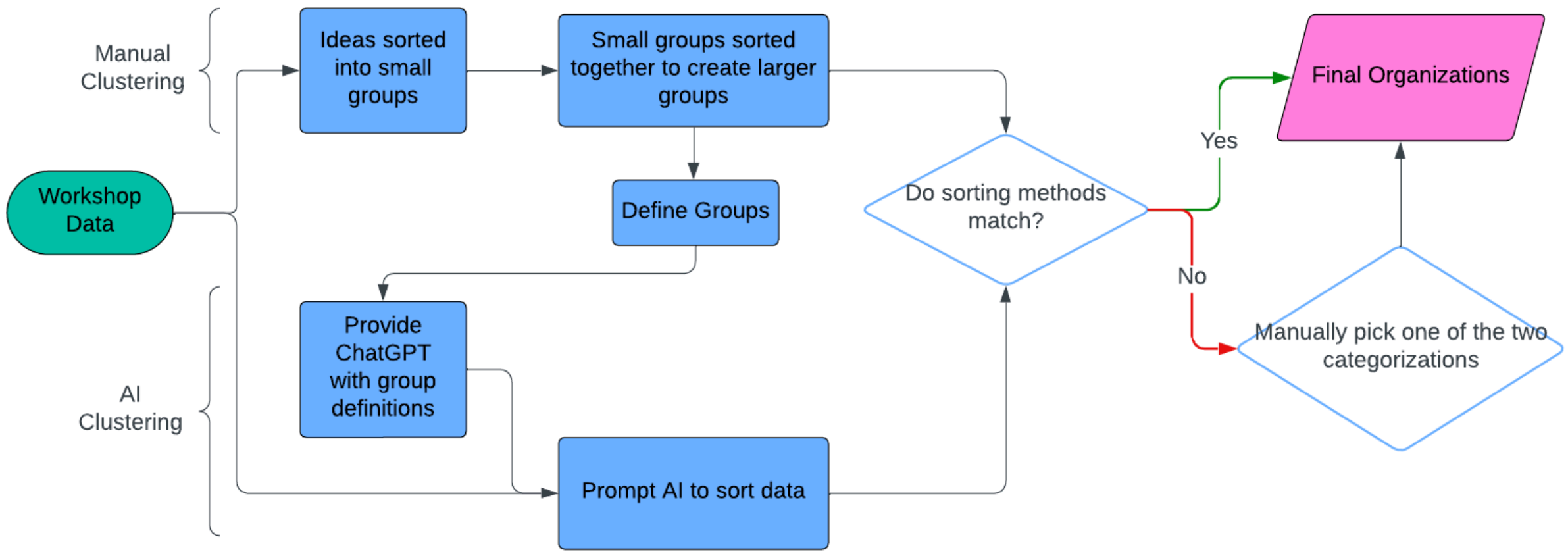

Figure 1. Flowchart depicting the idea clustering process involving human coders and a large language model.

Using this process, ideas were clustered into the groups described in Table 3 below. For the idea categories, a three-step approach was taken to quantify significance and popularity: 1) total votes per category, 2) normalized votes (votes divided by the number of ideas), and 3) a hybrid "value score" combining the first two approaches. A conditional analysis of category distribution across map regions was also conducted to understand where specific categories were most valued.

Sentiment and emotion analysis of workbook responses was done using the VADER Sentiment Analyzer [53] and Hugging Face Emotion [54]. Sentiment scores were averaged across sections, and emotions were classified by section, with separate analysis for students and community members. The results of these analyses are presented in the Findings section.

Table 3. Categorization of design ideas generated by the workshop participants.

| Idea category | Examples |
|---|---|
| **Community engagement**: Includes ideas pertaining to how the community interacts with the power plant. | Build a library around the reactor, Nuclear Science Museum, and Community stores within the plant |
| **Aesthetics**: Includes ideas related to appearance and sensory appeal. | Brightly colored and vibrant, Blends with the environment, Carved into a mountain, Music |
| **Environmental Protection and safety**: Related to safety for the environment and people. | Offset carbon emissions, Dispose waste safely, Plants to capture carbon |
| **Economically Beneficial**: Ideas about financial feasibility and economic benefits. | Reduce energy cost for consumers, Tax breaks for nearby residents, Bring investor money to the community |
| **Communication and ethics**: Focus on communication and ethical practices | Community-led board for future plans, Informational websites and pamphlets, Respect land rights |
| **Functionality / technical**: Ideas related to the facility's functionality. | Efficient grid connection, Partially underground design, Huron River for cooling |
| **Working environment**: Ideas about the workplace for employees. | Nightclub for workers, Fair wages and reasonable hours |
| **Unsorted**: Ambiguous or unrelated ideas | Homer Simpson |

# Results

The sections that follow describe the major findings from the community design workshop. We begin by introducing the values each of the ten teams identified as being significant for their work. We then describe the questions and concerns about fusion energy raised by the participants, the criteria chosen for making design decisions about the fusion energy facility, summary of the ideas generated, an analysis of participant sentiment and emotion, and a presentation of the facility design concepts created by each of the teams – as well as design themes we deduced from analyzing the design concepts. We close the results section by sharing the participants' reflections on working together to imagine and design hypothetical fusion energy facilities.

# Values

At the start of the workshop, each of the ten teams identified the values that mattered to them and which they expected would inform their design work together over the course of the day. Figure 9 shows the values across the ten teams. Values having a high frequency of occurrence within and across teams are shown in color. Notably, ‘integrity’ and ‘respect’ had the highest frequencies of occurrence across the teams, with all but three teams identifying these values as being important to them. Other values that were also noted across the teams included ‘hard work’, ‘health’, ‘equity’, and ‘family’. Two teams also identified ‘religion’ as a significant value. The shared values across the teams are notable, as each team independently identified values that were significant for its members. Therefore, the pattern of having both shared and unique values across teams suggests that teams also have shared and unique preferences, which, notably, translated into shared and unique design choices.

| TEAM 1 |
|---|
| INTEGRITY |
| HONESTY |
| RESPECT |
| SELF-PRESERVATION |
| HEALTH |
| COMMUNICATION |

| TEAM 2 |
|---|
| HONESTY |
| EAGERNESS TO LEARN |
| EMPATHY |
| GROWTH |
| DEDICATION |
| HELPFUL |

| TEAM 3 |
|---|
| HONESTY |
| HONESTY |
| HARD WORK |
| SAFETY |
| WISDOM |
| PASSION |
| DIPLOMACY |

| TEAM 4 |
|---|
| INTEGRITY |
| TRANSPARENCY/HONESTY |
| HARD-WORK |
| HARD WORK |
| RESPECT |
| RESPECT |
| EQUITY |
| PRODUCTIVITY |
| BE OPEN-MINDED! |
| THINK OUTSIDE THE BOX |
| CREATIVITY |

| TEAM 5 |
|---|
| RESPECT |
| RELIGION |
| THOUGHTFUL OF OTHERS AROUND AT ANY GIVEN TIME. |
| INCLUSION |
| CREATIVITY |

| TEAM 6 |
|---|
| TRUST |
| INTEGRITY |
| HARD WORK |
| TRUE INCLUSION |
| EMPATHY |
| SHARING KNOWLEDGE & RESOURCES NO GATE KEEPING! |

| TEAM 7 |
|---|
| TRUST |
| INTEGRITY |
| HARD WORK |
| RESPECT |
| RESPECT |
| RESPECT |
| HEALTH |
| HEALTH |
| HEALTH |
| EQUITY |
| FAMILY |
| FAMILY |
| STRUCTURE |
| COMPASSION |

| TEAM 8 |
|---|
| RESPECT |
| EQUITY |
| MORALITY/ETHICS |
| FAMILY |
| POSITIVE ECONOMIC IMPACT ON THE COMMUNITY |
| CURIOSITY |

| TEAM 9 |
|---|
| HONESTY/ INTEGRITY |
| HARD WORK TOWARD A PURPOSE |
| RESPECT |
| FAITH/RELIGION |
| RELIGION |
| LOYALTY |
| KINDNESS |
| NOT BEING ASSUMPTIVE |

| TEAM 10 |
|---|
| RESPECT |
| INDIVIDUALITY |
| CHALLENGE STATUS QUO |
| AUTHENTICITY |
| COLLABORATION |

| REOCCURING VALUES |
|---|
| INTEGRIY/TRUST |
| HARD WORK |
| RESPECT |
| HEALTH |
| EQUITY |
| FAMILY |
| RELIGION |

Figure 2. Values identified across the ten design teams. Recurring values are shown in color, with values shown in white being unique to each table.

## Questions and Concerns about fusion energy

As noted earlier, after being introduced to fusion energy, participants were allowed to ask questions and share their observations. These questions were submitted anonymously via Mentimeter and are shown below in Figure 3. Broadly, the questions and observations related to the challenges associated with commercializing fusion ("it seems like science fiction" and "very difficult to achieve"); the safety of fusion systems ("can a meltdown occur in a fusion reactor?"); the expected environmental impacts, including the management of fusion waste; expected lifespan of fusion energy systems ("what is the lifespan of a fusion reactor?"); and questions about the relative benefits of fusion energy compared to other energy sources. Some participants emphasized their excitement about fusion as a possible sustainable source of energy. Each of these questions and comments was addressed by the workshop facilitators in the larger group before lunch. Discussions on these questions continued within the individual teams during lunch, and the remaining questions were addressed by the students in the course, with some tables asking for further assistance from the course instructors to address the questions raised by the community participants.

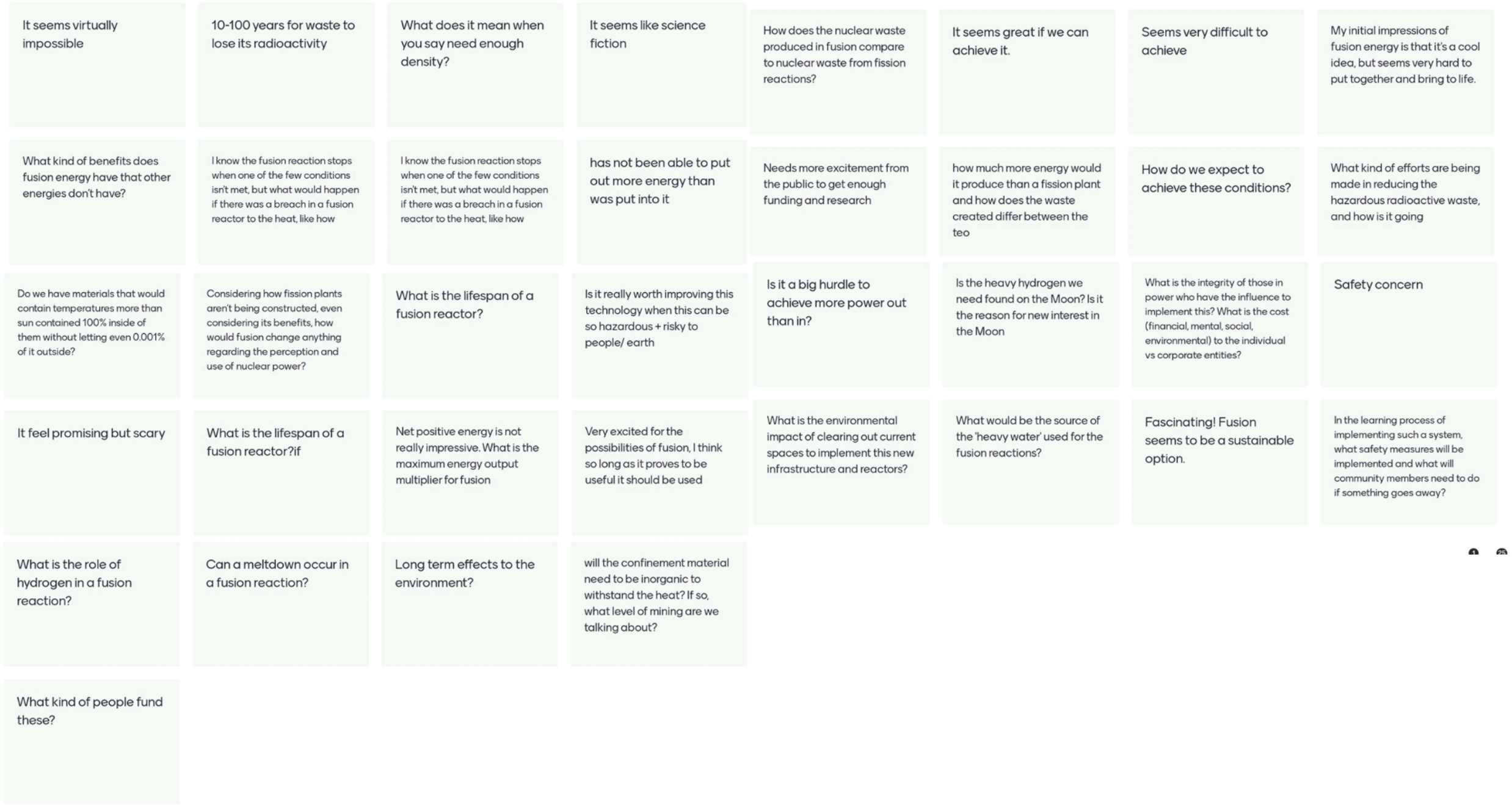

Figure 3. Concerns and questions about fusion energy generated by workshop participants using Mentimeter.

## Idea maps

In the next phase of the workshop, participants embarked on ideation and divergent thinking within their teams. Each participant was tasked with generating six unique ideas within six categories: technical, socioeconomic, environmental, community, aesthetic, and wildcard.

Participants were then asked to place these ideas on a map having four regions representing the community, the perimeter of the facility, the area for accessing the facility, and the fusion energy facility itself. Due to fusion energy concepts being in the relatively early stages of development and the sizes of each of these remaining unknown or undecided, we did not designate the size of these regions for the purposes of this workshop. Figure 4 shows an example of a completed idea map by one of the teams.

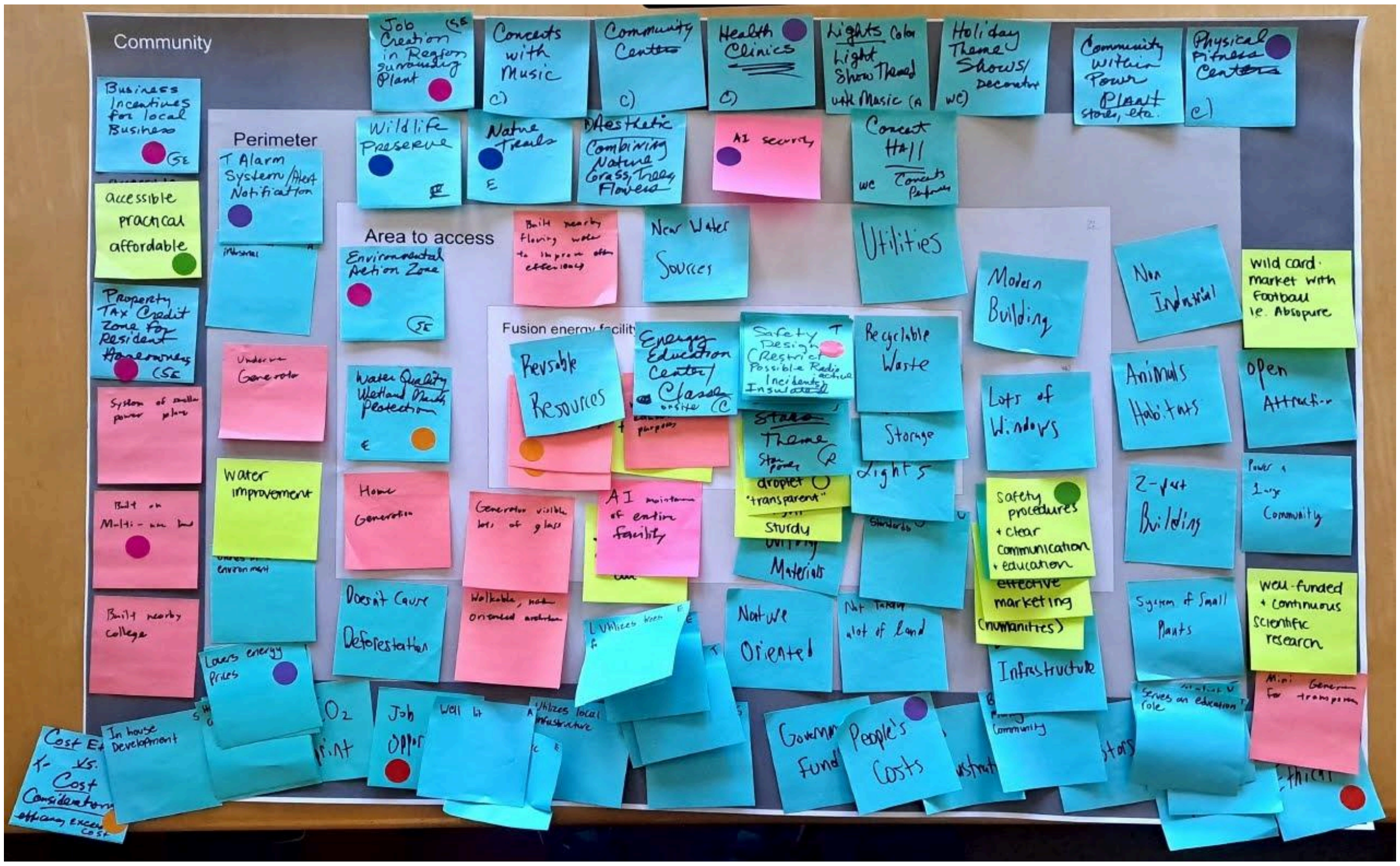

Figure 4. An example of an idea map completed by one of the teams. The map features four distinct areas of the energy facility, though not sized to scale. Ideas were placed in relevant locations on the map.

## Criteria

Following the phase of divergent thinking and ideation, participants entered a phase of convergent thinking during which they were tasked with determining decision-making criteria and downselecting from among their ideas. Participants on each team first individually identified decision-making criteria and then voted on the most significant, with each participant having the opportunity to vote for three criteria. These criteria, including those receiving the most votes, are shown in Figure 5 below. As with the values identified by the teams, there was a mix of shared and unique criteria across the teams. Safety, prioritizing the needs of the community, positive economic impact, cost effectiveness of the facility, creation of employment opportunities, protection of the environment, and low maintenance needs for the facility are some examples of shared criteria across the teams. Some examples of unique ideas include "makes community excited about the facility", "aesthetically pleasing", and not wasting the existing infrastructure and workforce, but reallocating them for the design and operation of the new facility are some examples of the unique criteria generated by some teams. Once identified, these criteria informed the teams' selection of ideas for their facility design concept.

| TEAM 1 | |
|---|---|
| MOST VOTED CRITERIA (3+) | VOTES |
| AFFORDABLE/ECONOMIC | 8 |
| ENVIRONMENTALLY FRIENDLY | 6 |
| SAFETY | 5 |
| LOW MAINTANCE | 3 |
| DON'T WASTE EXISTING INFRA/WORKFORCE; REALLOCATE EFFECTIVELY | 3 |

| TEAM 2 | |
|---|---|
| MOST VOTED CRITERIA (3+) | VOTES |
| HEALTH AND SAFETY | 6 |
| COST BENEFIT | 6 |
| BENEFITS ENTIRE COMMUNITY | 5 |
| JOB POSITIONS | 4 |
| EDUCATIVE | 3 |

| TEAM 3 | |
|---|---|
| MOST VOTED CRITERIA (3+) | VOTES |
| AFFORDABLE/ECONOMIC | 4 |
| ENVIRONMENTALLY FRIENDLY | 3 |
| SAFETY | 3 |
| LOW MAINTANCE | 3 |
| DON'T WASTE EXISTING INFRA/WORKFORCE; REALLOCATE EFFECTIVELY | 2 |

| TEAM 4 | |
|---|---|
| MOST VOTED CRITERIA (4+) | VOTES |
| SECURITY | 5 |
| ENVIRONMENTAL BENEFIT TO ECOSYSTEM (NET-ZERO CARBON, ETC.) | 4 |
| AFFORDABILITY | 4 |
| PRODUCTIVE/EFFECTIVE | 4 |
| HEALTH | 4 |

| TEAM 5 | |
|---|---|
| MOST VOTED CRITERIA (1+) | VOTES |
| PROTECT ENVIRONMENT | 2 |
| COST EFFECTIVENESS FOR ALL | 1 |

| TEAM 6 | |
|---|---|
| MOST VOTED CRITERIA (3+) | VOTES |
| POSITIVE ENVIROMENT IMPACT | 10 |
| INVOLVES COMMUNITY IN DESIGN DECISIONS | 5 |
| POSITIVE SOCIO-ECONOMIC IMPACT WITH CREATIVE GOOD LONG LASTING JOBS IN ITS COMMUNITY. | 5 |
| MAKES THE COMMUNITY EXICTED ABOUT FACILITY | 3 |
| BENEFITS THE MOST AMOUNT OF PEOPLE | 3 |

| TEAM 7 | |
|---|---|
| MOST VOTED CRITERIA (3+) | VOTES |
| ECONOMICALLY POSITIVE | 7 |
| SAFE FOR ENVIRONMENT | 5 |
| NEW CAREERS/JOBS | 4 |
| AESTHETICALLY PLEASING | 3 |
| COMMUNITY ENGAGEMENT | 3 |

| TEAM 8 | |
|---|---|
| MOST VOTED CRITERIA (3+) | VOTES |
| SUSTAINABLE (ENVIRONMENTALLY & OTHERWISE) | 5 |
| SHOULD ULTIMATELY MAKE ENERGY MORE AFFORDABLE | 5 |
| POSITIVE IMPACT ON THE LIVES OF AS MANY PEOPLE POSSIBLE, W/O NEGLECTING OTHERS | 4 |
| ENCOURAGING THE USE OF FUSION ENERGY INSTEAD OF MAKING IT BARELY ACCEPTABLE | 3 |
| SOCIOECONOMICALLY BENEFICIAL | 3 |

| TEAM 9 | |
|---|---|
| MOST VOTED CRITERIA (3+) | VOTES |
| PRIORITIZE THE NEEDS, JOBS, AND SATISFACTION OF THE PEOPLE | 4 |
| COST BENEFIT ANALYSIS IS MAINLY POSITIVE | 4 |
| BALANCE OF ENERGY BETWEEN GOVERNMENTAL, AND PRIVATE COMPANY - FOR GOVERNMENT AND PRIVATE CAN HAVE FULL CONTROL OF THE SOURCE OF ENERGY | 3 |
| SUPPORT/BUILD THE ECONOMY | 3 |

| TEAM 10 | |
|---|---|
| MOST VOTED CRITERIA (3+) | VOTES |
| KEEPING THINGS COST EFFICIENT, YET HAVING MAXIMUM ENERGY PROFICIENCY. | 6 |
| SUSTAINABLE/ENVIRONMENTALLY FRIENDLY | 5 |
| MORE ENERGY EFFICIENT THAN OTHER CURRENT ENERGY SOURCES | 3 |
| MEETS COMMUNITY ENERGY NEEDS | 3 |
| INCENTIVIZES COMMUNITY | 3 |

Figure 5. Decision-making criteria were identified across the ten teams

The distribution of participant votes across the design categories is shown in Figure 6. Notably, the three categories of ideas receiving the most votes across all teams were environmental protection and safety, economically beneficial, and community engagement categories. The technical, aesthetic, and community engagement categories received roughly a similar number of votes, with the working environment being the smallest substantive idea category in terms of votes.

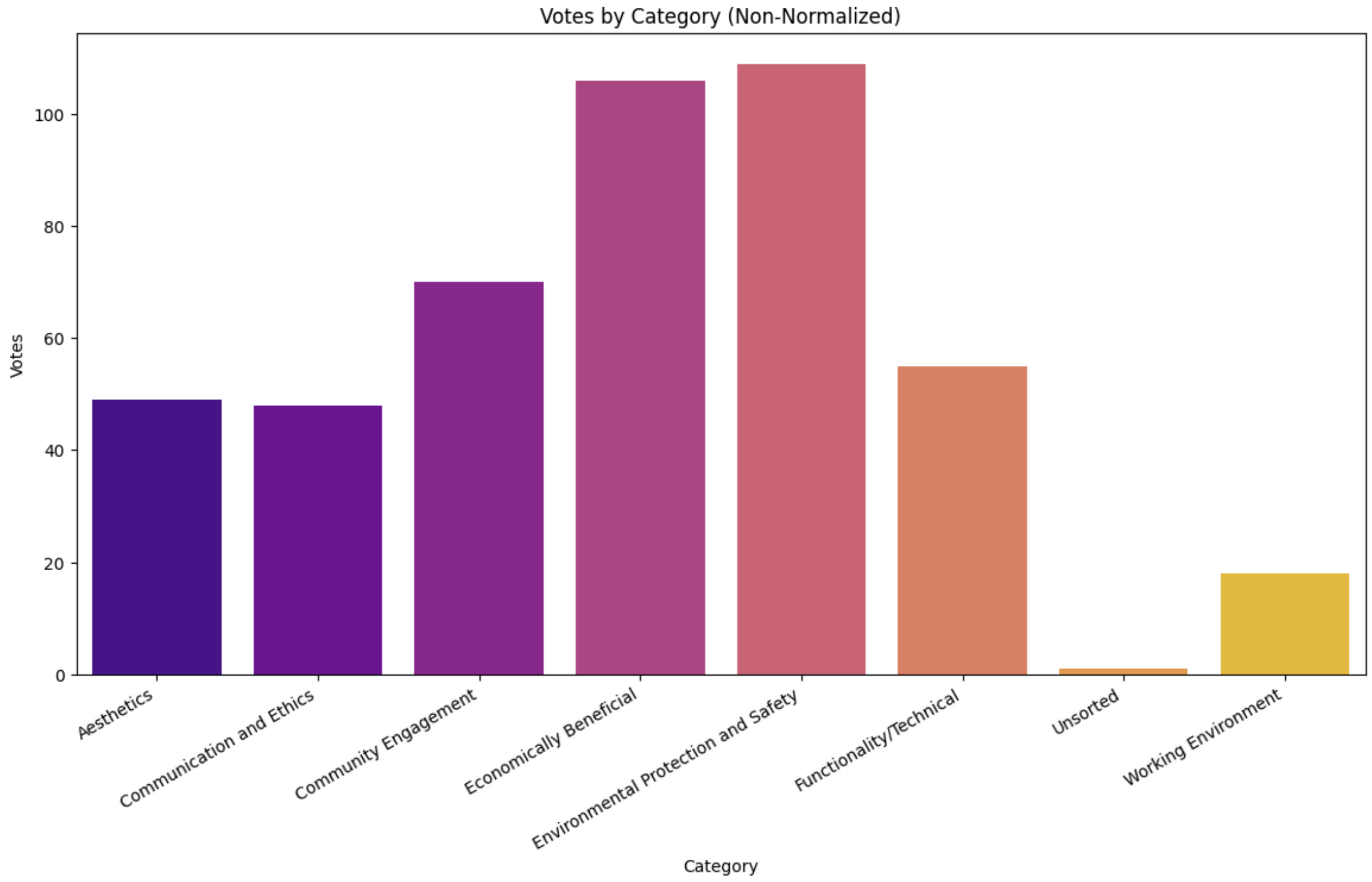

Figure 6. Distribution of participant votes across the design idea categories. Environmental protection and economic benefits for the community were primary concerns.

The ideas were further sorted to show how each of the idea categories shown in Figure 6 occurred in the four different regions of the map. Here, unsurprisingly, the fusion energy facility region showed a high concentration of ideas focused on the technical and functional aspects of fusion energy. Interestingly, ideas related to aesthetics, environmental protection, and safety also showed a high occurrence in this region. Each of the other three areas showed a reduced emphasis on technical and functional ideas. Aesthetics remained a significant idea category for accessing the facility and its perimeter, but declined in importance for the area designating the community. This finding might suggest that while the participants are interested in careful and deliberate design of the aesthetics of the fusion energy facility, they may be less interested in altering the aesthetics of the host community itself. The three most significant idea categories in the ‘community’ area were the economically beneficial, communication and ethics, and community engagement categories, which reinforced the intent of the participants for the facility to have a positive economic impact in the host community and for the community to remain engaged in the design and operation of the facility.

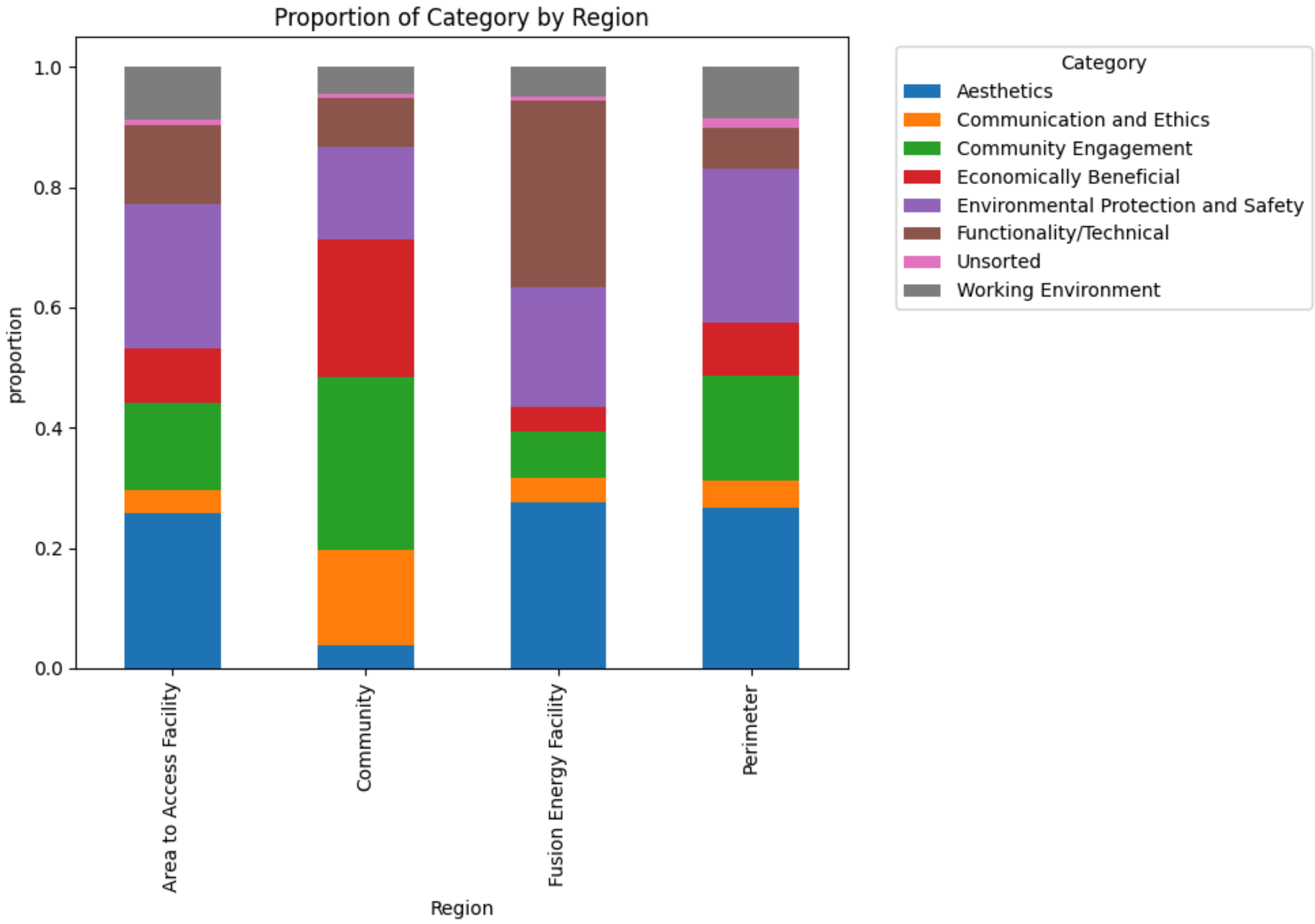

Figure 7. Distribution and significance of idea categories across the facility map

## Emotion and Sentiment

The VADER Sentiment Intensity Analyzer was used to quantify the sentiment on individual pages of workshop participants' workbooks. The content of each workbook page is briefly described in Table 4. The bulk of participants' written responses were documented on pages 6, 7, and 8. Page 6 included participants' views on fusion energy, including what they were confused about and had learned at the workshop. Page 7 was a space for them to brainstorm ideas for their ideal fusion energy facility, and page 8 included their criteria for how they would select their most important ideas from page 7.

Table 4. Description of workbook page content by page number

| Page(s) | Content |
|---|---|
| 1-3 | Introductory (included welcome statement and agenda) |
| 4 | Community agreements |
| 5 | Self introductions |
| 6 | Key takeaways and new questions about nuclear energy |
| 7 | Design ideas for an ideal nuclear reactor. Participants were asked to provide at least one idea for each of these six categories: technical, socioeconomic, environmental, community, aesthetic, and wild card. Wild card was for anything that did not fit in the other categories. |
| 8 | Criteria that participants would use to select their most important ideas from page 7. |
| 9-10 | Notes |
| 11 | Two words to describe participants' experience with the workshop. |

The sentiments were quantified with negative, neutral, and positive scores between 0 and 1 and a singular compound score. The sentiment scores for all responses were averaged per page to get a conditional sentiment score per page, shown in the figure to the right. We can see that all pages, barring the first three, which have little to no participant responses, have an overall positive connotation as indicated by a compound score greater than 0.2. Certain pages, such as those relating to fusion or notetaking, had more neutral sentiment, while almost none had negative sentiment. Because participants filled out the workbook pages sequentially as the workshop progressed, the participant sentiment, as analyzed across the workbook pages, can also be used as a proxy for the sentiment over the course of the workshop. This analysis suggests an overall neutral to positive sentiment.

In addition to participant sentiment, we also analyzed participant emotion across the workbook pages, separating the student workbook responses from those of the community participants. Emotion categories as analyzed included neutral, fear, joy, sadness, surprise, disgust, and anger. From Figure 9 below, we see that both students and participants recorded largely neutral sentiment on the initial pages, with a rise in surprise expressed by students on the later pages of the workbook and a rise in joy expressed by community participants on the later pages when the workshop transitioned from team formation and learning about fusion to designing the fusion energy facility. For example, one student wrote that "[m]any new perspectives & views from community members led to us in turn being more creative in and of ourselves." Other students described the experience as it was unfolding as "interesting", "surprising", and "insightful". The community members wrote about feeling "comfortable and positive about community centered design". Others wrote about feeling hopeful,  invigorated ( "I feel hopeful and invigorated") and educated ("I feel educated and hopeful about nuclear [fusion]".)

Table 4. Distribution of participant sentiment across the workbook pages

| Page | neg | neu | pos | compound |
|---|---|---|---|---|
| 1 | 0.000 | 1.000 | 0.000 | 0.000 |
| 2 | 0.000 | 1.000 | 0.000 | 0.000 |
| 3 | 0.000 | 1.000 | 0.000 | 0.000 |
| 4 | 0.032 | 0.722 | 0.246 | 0.402 |
| 5 | 0.020 | 0.740 | 0.239 | 0.681 |
| 6 | 0.045 | 0.841 | 0.114 | 0.268 |
| 7 | 0.052 | 0.803 | 0.146 | 0.519 |
| 8 | 0.036 | 0.671 | 0.292 | 0.545 |
| 9 | 0.003 | 0.734 | 0.262 | 0.460 |
| 10 | 0.026 | 0.799 | 0.175 | 0.397 |
| 11 | 0.021 | 0.457 | 0.522 | 0.517 |

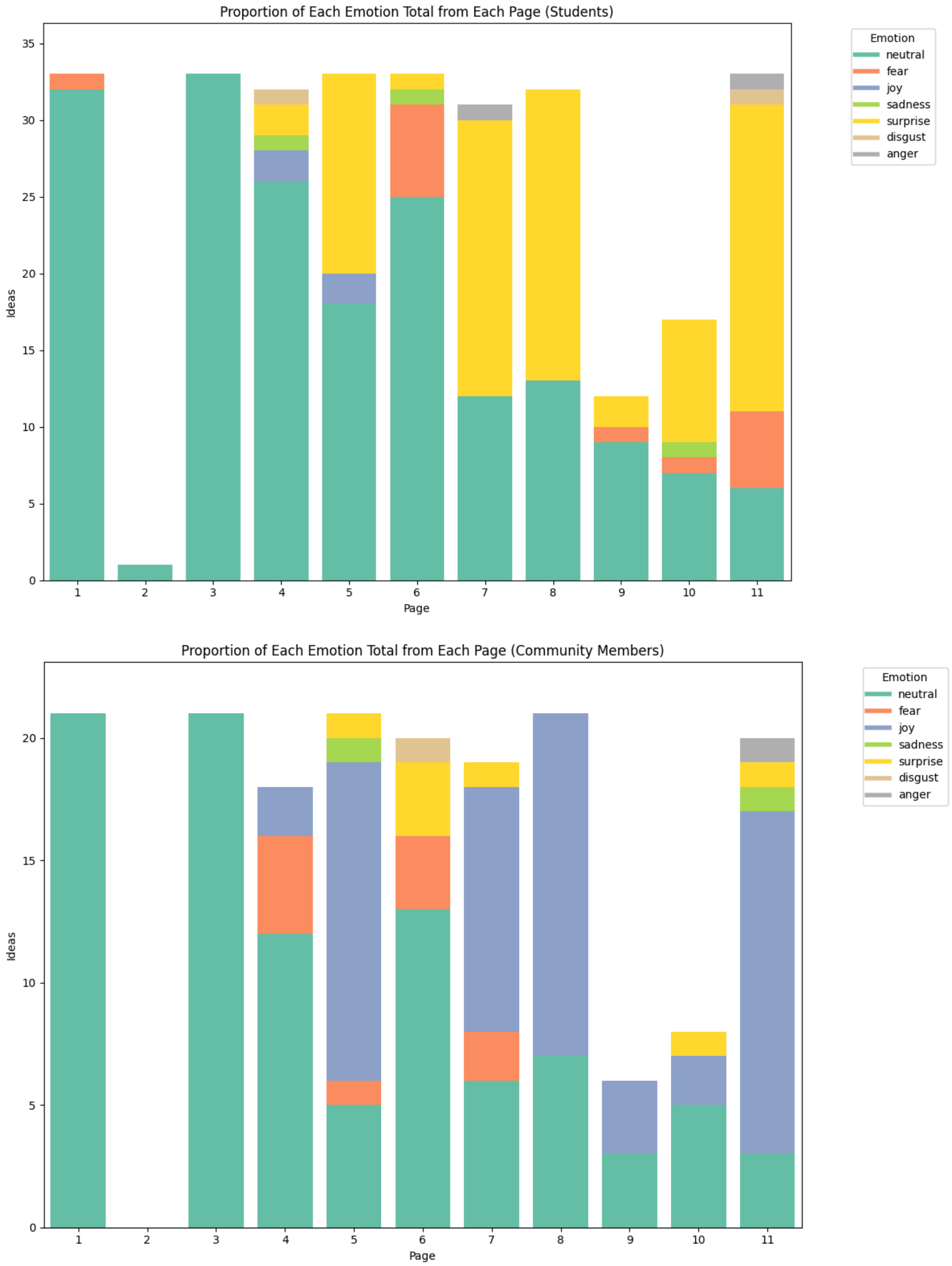

Figure 8: Distribution of participant emotion across the workbook pages. Neutral or surprised feelings were predominant among students, whether neutral or joyful feelings were predominant among community members.

## Facility design concepts

Once the teams had downselected a subset of ideas from the larger pool of ideas initially developed, they textually described their design concepts and developed visual representations or prototypes of their facility designs using generative AI. Students and community participants were free to use generative AI models of their choosing. Some examples of models used included Canva's AI image generator, Adobe Firefly, and DALL-E2 E2. These facility design concepts and the images that represent them are reflective of each team's values, design criteria, collective hopes, and concerns for their communities in the context of energy infrastructure. In our analysis of the facility design images generated by the workshop participants, we observed a five salient themes that are highlighted in the sections below.

### Connecting the design of the facility to the history and legacy of the community

One particularly salient theme was a desire to connect the history and legacy of the community to the design of the facility. For example, one of the teams featured community participants from Detroit. This team, which called itself 'From Cars to Stars', sited their facility in the Detroit, MI area and believed the energy facility should be reflective of Detroit's industrial history of car manufacturing, including its rise, fall, and ongoing renewal. Their design included plans to repurpose abandoned buildings, use them in their imagined facility, and more broadly to turn Detroit into a manufacturing hub for a new energy industry. Another team similarly echoed the expectation that an initial fusion facility could start the creation of a local industry. Participants from this team observed that perhaps "fusion funds itself." They expected profits from an initial fusion power plant could be "used to build more plants & educate more people." Another team noted the importance of balancing control and power between private companies and the government. They speculated about the possibility of creating a community-owned fusion power plant that would "support [and] build the local economy with a positive cost-benefit ratio." In this manner, several teams' design choices were informed by the history of Southeast Michigan and themes of economic renewal and industrial revival.

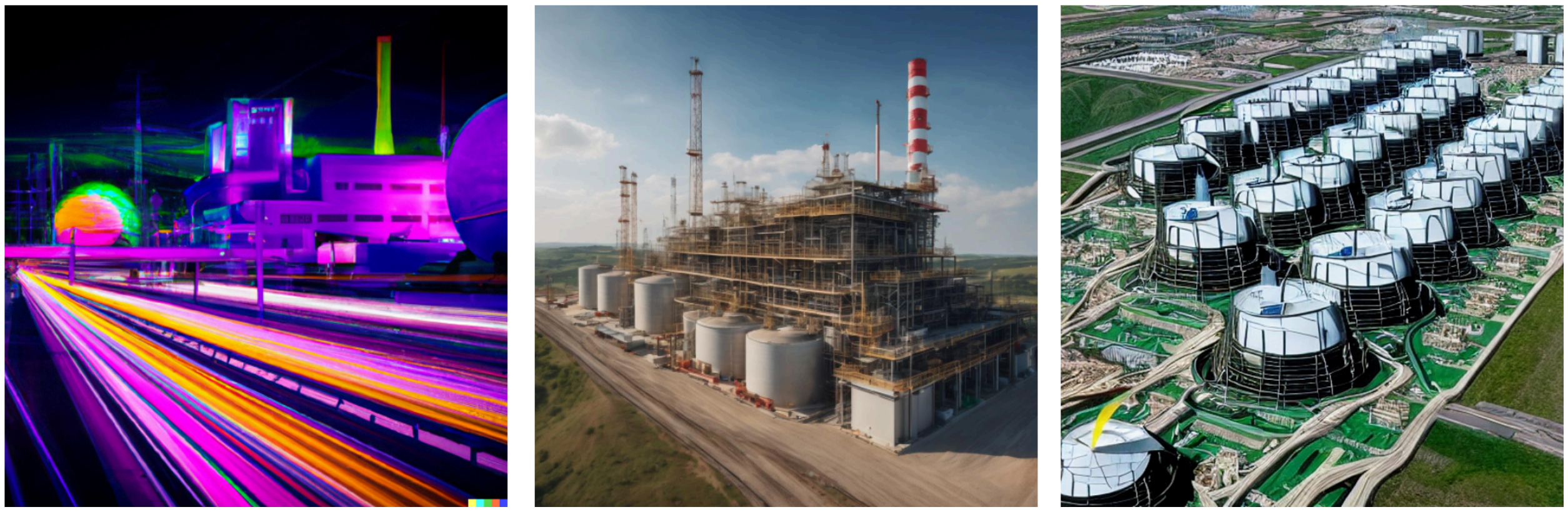

Figure 9. From left to right, the first image shows a facility located in Detroit, MI, and connected to communities via high-speed rail. In the center and left are "community-owned" fusion energy systems.

### Respect for Nature

Another salient theme that emerged in the designs was the respect and care for nature. This preference tended to manifest in several different and unique ways across the teams. Several teams created designs that show fusion energy facilities that are embedded in (rather than disruptive of) natural environments like open fields, by rivers, or even forested locations.

Other teams were particularly concerned with the environmental impacts a fusion energy facility might have on local communities. In these cases, to minimize impacts on the ground, one team even imagined that energy facility workers could move from one part of the system to another via a series of zip lines. Rather than concrete pavement, this team also chose to incorporate nature trails into their facility design. Some teams also expressed their concern for the environment and the community by incorporating specific stipulations about the management of fusion waste into their facility concepts. One team noted that it would be important to "repurpose waste from the facility," while another team prioritized "low waste production" and ensuring that "waste is safely contained and recycled." Others noted the importance of building "environmental impact awareness" and "limit[ing] impact on the environment.

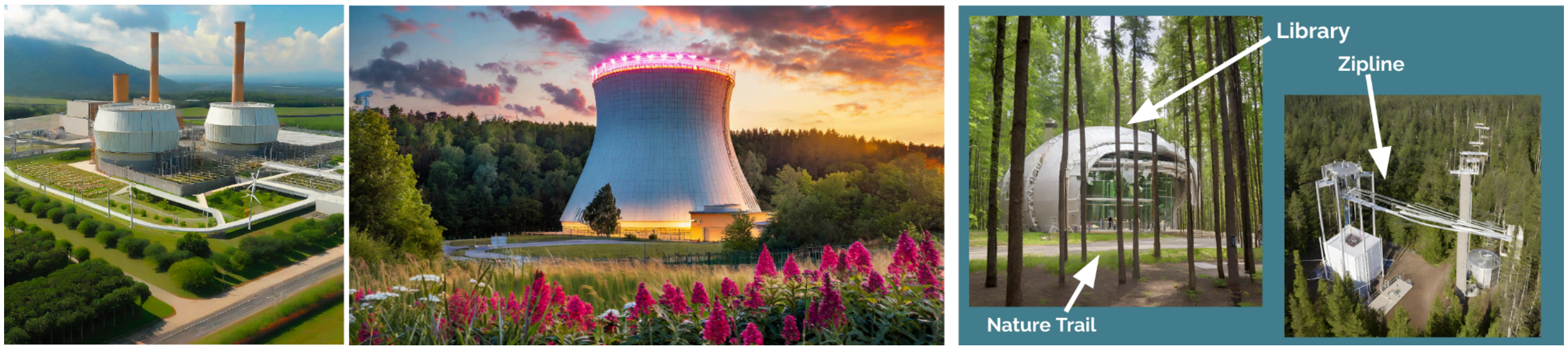

Figure 10. Each of these images was generated by teams that centered "nature" and its preservation in their design process. Third from the right is a team that integrated ziplines and nature trails into their facility design.

### Care for Workers

A third salient theme evident in the teams' designs was care for the employees who would work at the fusion energy facility. This care for workers manifested itself in the form of prioritizing their health and safety, ensuring that the facility created well-paying jobs, and that support services (such as child care) were provided to remove barriers to work.

The emphasis on worker health and safety can likely be tied to well-known accidents at nuclear fission facilities and lingering public fear about nuclear energy more broadly. At the same time, these teams' commitment to worker safety also illustrates the closeness or sense of community that characterizes many regions that may be good candidates for host fusion facilities in the future. In other words, these teams see the fusion workers as "us" and  "one of us" and want to be sure they are designing and inviting in a facility that is safe for everyone. Many teams specifically wrote about "community jobs" and "equitable jobs." One team enumerated "amenities" for workers they hoped the facility would provide, including "[r]easonable work hours, pays well, gym, sauna, free healthcare, and transportation." Some teams also wanted the facility to offer internships or work-study programs so that people in the community could be trained in its use and maintenance (rather than simply hiring people from elsewhere).

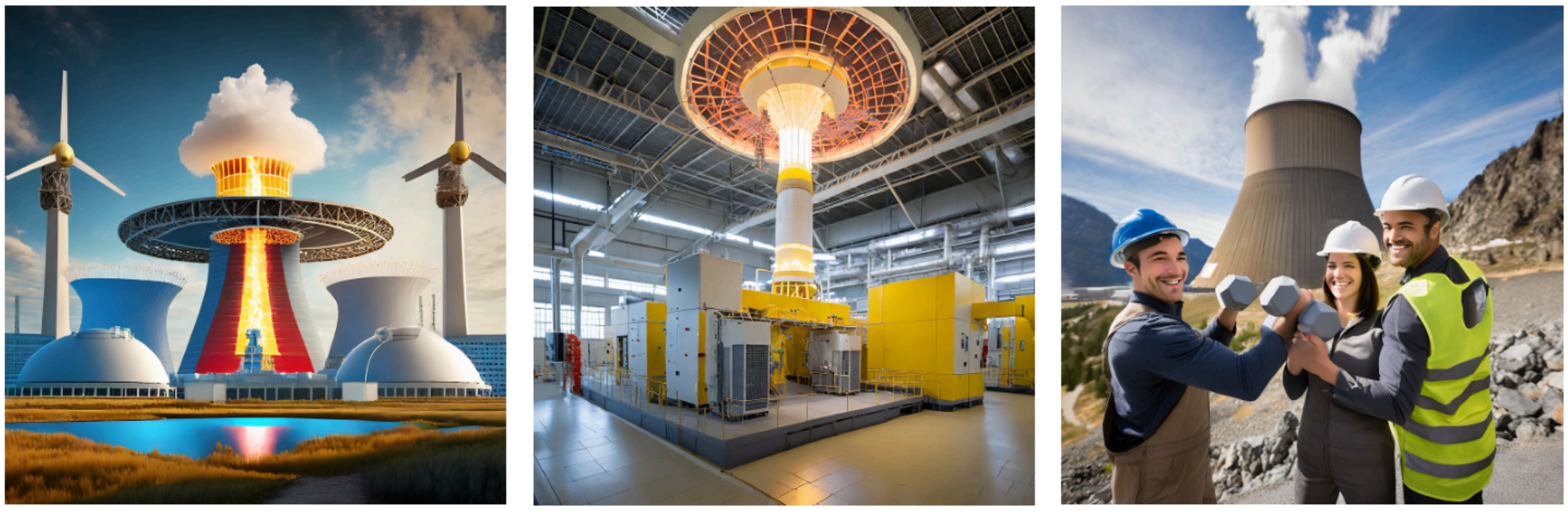

Figure 11. Facilities are designed with mutual worker/community benefits and safety in mind.

### Transparency and access to the facility

A fourth motivating theme for several teams was to create an energy facility that was ethical and transparent in its construction, operation, and communication with community members. The emphasis on transparency and implementing transparency in design appears, for some teams, to be the direct result of the values selected earlier in the day. One team interpreted this idea of transparency quite literally but creating a facility with all exterior walls being transparent. There were other, unique interpretations of transparency as well, with one team calling for the creation of an interactive map showing energy usage at the facility. This idea arises from the fact that energy facilities, though net producers of electricity, also require electricity for operation. Other teams called for the availability of “publicly accessible information” and “publication of reports to the public for widespread information.”

Similarly, these teams were interested in designing a facility that could be accessible, at least in part, to community members so they could observe some of its operations firsthand and feel as though the facility was part of the community rather than looming or otherwise separate from it. In this vein, one team called for the design of a community center at the facility and proposed that such a center would feature a museum, childcare, as well as a “winter warming station”. Similarly, another team also proposed the creation of a museum and offering tours of the facility. Many teams also expressed a keen interest in integrating public art projects into the design of the facility.

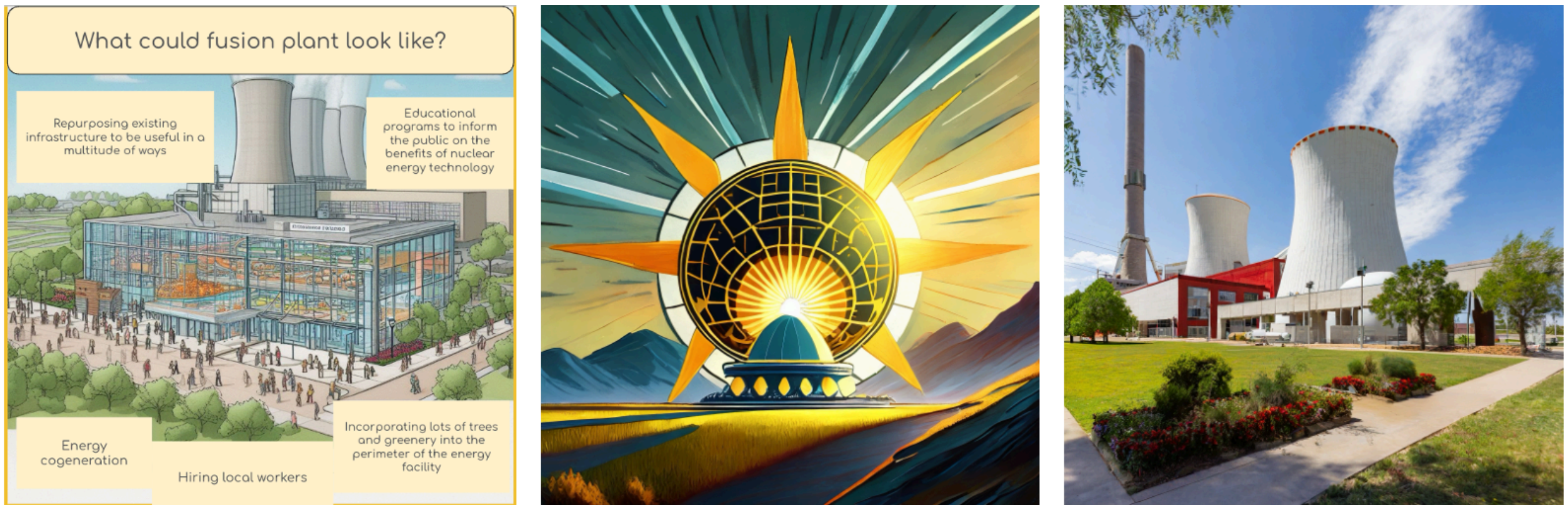

Figure 12. Facilities designed with transparency as a central value and/or design criterion. Teams wanted the energy facilities to be visually appealing, open to the public, and honest in their communications.

### Health and safety of the community

Another salient theme, though not easily illustrated in pictures, but captured in the textual descriptions developed by the teams concerned, the health and safety of the community hosting the fusion energy facility. While, as we have seen above, some teams had a preference for embedding the energy facility into the community, others called for “plac[ing] the facility far from people” as a way to ensure “safety for [the] community.” Other teams articulated this concern by calling for “safety, security, and public health measures” to be implemented at the facility. While the teams did not propose concrete design choices (other than in one case, placing the facility far from people) to ensure safety, it was clear from the textual facility descriptions that the health and safety of the community was an overwhelming priority for the teams. Many of the health and safety-oriented teams, also called for transparency of decision-making as described above, which suggests that these teams had a preference for the community to be able to directly verify the safe operation of the fusion energy facility.

## Participant reflections

Participants at the workshop were able to reflect on their experiences of working together in several ways. They wrote about their individual experiences in their workbooks, as well as shared two words that summed up their experience using Mentimeter. They also reflected on their shared experiences in their presentations to the whole group.

Twenty-four responses recorded using Mentimeter were sorted into three sentiment categories: positive, neutral, and negative. Nineteen of the responses were positive and five were neutral. We did not record any negative responses via Mentimeter. It is possible that some participants, although having negative or critical responses, may have chosen not to voice them. However, given the ability to respond anonymously on Mentimeter, we think it unlikely that the positive responses are performative and that the missing responses fall in the negative category. Among the ‘positive’ coded responses, participants expressed feeling “hopeful”, “optimistic”, “inspired”, “empowered”, “creative”, and “comfortable”. The ‘neutral’ coded responses included one participant finding the experience “pretty okay,” whereas others expressed feeling “curious” and “engaged” and referred to the workshop as “multifacetedly informative”.

Table 5. Individual two-word summary responses on the community-engaged design experience shared by participants

| Positive | Neutral | Negative |
|---|---|---|
| **(19 responses)** | **(5 responses)** | **(0 responses)** |
| Nuclear, joy | Interesting, exploration | |
| Surprising, insightful | pretty okay | |
| Hopeful and optimistic | ideating, collaborating | |
| Informed, familiar | curious and engaged | |
| Inspired, empowered | multifacetedly informative | |
| encouraged, hopeful | | |
| inspired, interesting | | |
| creative, fresh | | |
| creative and interested | | |
| englightened & intrigued | | |
| encouraged, hopeful | | |
| community, excited | | |
| fascinating, impactful | | |
| excited, creative | | |
| comfortable and positive | | |
| smooth and interactive | | |
| inspired and hopeful | | |
| heartened, optimistic | | |
| hopeful and informed | | |

Participants shared reflections centered on three main themes: ideation and managing ideas in the design process, ease of working as a team, and mutual learning.

**Managing ideas in the design process**

On the theme of ideation and managing ideas in the design process, some teams found it easy to think divergently and come up with new ideas, but difficult to manage the large number of ideas that emerged during the divergent thinking exercise. One team noted that, "It was easy to come up with ideas, it was difficult managing all the ideas we did have." Another noted that although they generated a large number of design ideas for the facility and "lots of diverse thought", they were able to ultimately generate consensus "on most ideas". This team stated that they "found there was often a lot of overlap in our ideas and many ways to connect them."

However, one team specifically noted that they found it challenging to develop a large number of unique ideas relating to the technical features of the facility. This observation may point to the need to dedicate more time to inform participants about fusion energy (or other technology in question) for future workshops, particularly when the outcomes are likely to shape real-world decisions and choices. As part of the course within which we held the participatory design workshop described in this paper, we also developed cross-sectional and facility-level VR models of ITER [55]. Community members could be invited to explore these VR models in the future to build their understanding of fusion energy. Such engagement could take place as a precursor to the participatory design workshop.

**Teamwork inspires creativity**

A critical learning from this process, however, is that working as a team made it possible for the participants to generate unique ideas and have a meaningful, shared experience. As one team noted, “Working as a team made it easy to generate ideas, as we combined the creativity of all our members.” This team was of the view that working together made it possible for them to “think like a kid” and “to be able to generate ideas without feeling judged.”

Across the teams, there appeared to be a consensus that the participants found it easy and even enjoyable to work with each other to design the hypothetical fusion energy facility. Many observed that they “worked well with the team”, that it was a “good experience to work as a team”. Other teams noted that while their members did not have a background in nuclear science or fusion energy, their passions and interests meaningfully informed design choices for the hypothetical energy facility intended for their community. As one team stated, “It was enjoyable. Everyone brought their own expertise and passions that, while they [may] not be directly about nuclear, they can be applied. It was good to learn how other people view a common subject.”

**Unique backgrounds facilitate mutual learning**

On the theme of mutual learning, many teams noted in their presentations that the experience helped them learn how a person’s background shapes their decision-making. As one team explained, “We all learned a lot about how where one comes from determines a person’s values.” Similarly, another team shared that they“ learned how even though we all came from different backgrounds as well as having a large age disparity, we all ultimately had the same core ideas.” In this vein, a team specifically noted that the variety of participant backgrounds that informed their decision-making in fact contributed to “sound results” they could have confidence in. In their shared reflections, many teams commented on how much they learned from one another and noted the value of strong communication practices (“Open communication optimizes the design process at all levels.”).

# Discussion

These results suggest at least four key takeaways regarding the value of participatory design as applied to fusion energy system development.

**Participatory design makes community preferences and concerns visible**

Our findings suggest that a participatory design process is an effective way to seek public perspectives and preferences even in the design of very early-stage technologies. Because concrete designs for fusion power plants that the public can examine and assess do not yet exist, we informed the participants about the science and engineering underlying fusion energy generation and used this knowledge as a starting point for designing a hypothetical facility. Often, when existing concrete designs exist, designers tend to fixate on these designs, finding it difficult to think divergently or explore creative ideas – a phenomenon referred to as ‘design fixation’[56,57]. In our case, the absence of concrete designs may have allowed the participants to think freely and make design choices that they value most. Indeed, we saw this kind of thinking in the decisions many teams made to integrate nature into the design of the facility, to design various forms of transparency, as well as the integration of community-facing aspects of the energy facility, including public art projects, museums, community centers, and facility tours. These are not typical features of energy facilities, and are therefore more likely to be indicative of the participants' preferences as opposed to being widely-held expectations of what an energy facility should look like. In short, a participatory design process, even when concerned with the design of a purely hypothetical facility, may be valuable in that it can generate design ideas and facility-wide concepts that could be implemented at a later date for a real-world facility. We therefore view a participatory and speculative design approach, such as the one implemented here, as a way to build community-sourced idea banks from which future technology developers can draw inspiration.

**Participatory design and achieving social license**

While we certainly acknowledge that many of the designs proposed above could not come to fruition exactly as imagined, we want to highlight the value of the workshop process (and knowledge-sharing) that informed these designs. Though many participants started with a somewhat skeptical perspectives on fusion (as seen in the questions and concerns raised about fusion early in the workshop process), they appear to have come away from the workshop feeling inspired to learn more about fusion energy, share their experiences with other people, and participate in future participatory design opportunities. Our findings suggest that the process of learning with and from each other and engaging in a creative design process to imagine a hypothetical fusion energy facility improved both sentiment and understanding. To be clear, we are not calling for an instrumental use of the participatory design process with the goal of building ‘acceptance’ for a technology. Instead, our findings suggest that participatory design processes can help build mutual trust among publics and experts, which can ultimately lead to social license. Finally, as fusion energy approaches commercialization and facility design changes from creating the hypothetical to the concrete, it will be important to examine and understand the impacts and possible changes on participant sentiment and willingness to participate.

**Valuing location-specific preferences**

The participants' location-specific knowledge and preferences highlight the need for location-specific energy solutions. Communities' unique preferences, histories, and cultures are vital to successful and enduring energy infrastructure design. As can be seen from the results, we observed a significant variation in community values, decision-making criteria, and design choices even within the specific region – Southeast Michigan – that has been the focus of our work. Understandably, there is a tension between standardizing fusion energy system designs and community-specific facility-level solutions. Fusion energy developers are likely to pursue design standardization for reasons of economic efficiency, safety, and even workforce training. However, a question worth considering in the relatively early stages of fusion energy system development is whether an identical fusion device can be integrated into a range of different facility types, which might be more community and context-specific. If such an approach is pursued, a corollary objective should be to design the fusion device itself to have a high level of sociotechnical readiness [58] such that the implications for safety, as well as socioeconomic and environmental impacts, are considered acceptable across a wide range of facility designs into which that fusion device could be integrated.

**Designing an energy facility from inside the community**
Most complex engineering systems are designed from the inside out, starting with the central technology (in this a fusion energy system – be it a tokamak, stellarator, IFE-based, or hybrid device) and building the secondary and tertiary systems around that technology and then, ultimately, the energy facility. In most instances, communities that host energy facilities are consulted not at all or in the very late stages of the facility development process, when community input can do little to meaningfully alter design choices. The model of decision making put forward in this study takes a different approach, which emphasizes not just the core technology as a central focus of the design process, but instead views the facility holistically, with design choices in many cases made in an *outside-in* manner such that considerations of socioeconomic, aesthetic, and environmental impact, shape design choices from the outset. Certainly, there are some challenges with pursuing such an approach to design for an early-stage technology still in the demonstration phase, across a wide range of possible concepts, with engineering efforts still largely focused on achieving important goals such as a sustained fusion reaction and net energy gain. While these are vital priorities, we conjecture that paying, if not equal, then at least significant attention to the facility level design choices – and making these choices in an integrated way across all hierarchical levels of the complex system is likely to ultimately produce designs which will earn social license.

# Future work

Moving forward, we have at least three specific aims in mind.

First, provided approved funding, we aim to continue hosting participatory design workshops in our first-year engineering course and continue to build connections with community members and local organizations whose members might be interested in participatory design. This model offers an opportunity to teach engineering designers how to use the participatory design process and also improve our workshop model to suit a variety of contexts.

Specifically, and as a second aim, we hope to bring the workshop model into international contexts to gather cross-cultural perspectives on fusion energy system design. In this case, we will partner with interested fusion developers to offer workshops in their local communities as a way to support their design process and open up dialogue between communities and emerging fusion companies. Future workshops will offer opportunities for participants to interact with VR models of fusion reactors mentioned above. Simultaneously, we are also developing an online generative-AI-based tool that allows users to share their visualizations of and reflections on fusion (as well as other clean energy facilities). A beta version of this tool has already been developed [59]. We also have developed an online learning platform, Global Fusion Forum (https://gff.fptz.org/), that invites users to learn fusion energy systems basics and also share their perspectives about its development and deployment. Through the workshops and the online tools, we seek to create (as noted above) a community-sourced idea bank for fusion energy facility designs which can inspire and inform real-world design choices made by fusion developers.

And finally, from this work, we aim to develop a “playbook” for participatory design of fusion energy systems that could be widely distributed and used in industry and education contexts. The playbook would offer a clear plan for engaging communities in participatory design while also providing flexibility or adaptability to specific contexts of use.

# Conclusion

Participatory design of fusion energy systems can be a valuable tool for improving communication, trust, and understanding between experts and publics involved in this process. The model we have developed shows promise as a tool to help achieve social license as the fusion energy moves toward grid-scale deployment. This process is replicable and scalable to industry and educational contexts, and at the same time, may make the historically contentious process of energy system siting an opportunity for collaboration and mutual learning.